\documentclass[12pt,preprint]{aastex}

\usepackage{graphicx}
\usepackage{epsfig}
\usepackage{times}
\usepackage{natbib}
\usepackage{url}
\usepackage{color}
\usepackage{amssymb,amsmath}
\usepackage{verbatim}

\newcommand{\an}{{Astron. Nachr.}}

\newcommand{\q}[1]{{\color{black}{#1}}}

\shorttitle{Writhe Of Kink-Unstable Flux Ropes}
\shortauthors{T\"or\"ok et al.}

\begin{document}

\title{The Evolution of Writhe in Kink-Unstable Flux Ropes and Erupting Filaments}

\author
{
T.~T\"{o}r\"{o}k       \altaffilmark{1},
B.~Kliem               \altaffilmark{2},
M.~A.~Berger           \altaffilmark{3},
M.~G.~Linton           \altaffilmark{4},
P.~D\'emoulin          \altaffilmark{5},
L.~van Driel-Gesztelyi \altaffilmark{6,5,7}
}
\altaffiltext{1}{Predictive Science Inc., 9990 Mesa Rim Rd., Ste 170, San Diego, CA 92121, USA}
\altaffiltext{2}{Institut f\"ur Physik und Astronomie, Universit\"at Potsdam, Karl-Liebknecht-Str. 24-25, 14476 Potsdam, Germany}
\altaffiltext{3}{University of Exeter, SECAM, Exeter, EX4 4QE, UK}
\altaffiltext{4}{U.S. Naval Research Lab, 4555 Overlook Ave., SW Washington, DC 20375, USA}
\altaffiltext{5}{LESIA, Observatoire de Paris, CNRS, UPMC, Univ. Paris Diderot, 5 place Jules Janssen, 92190 Meudon, France}
\altaffiltext{6}{University College London, Mullard Space Science Laboratory, Holmbury St Mary, Dorking, Surrey, RH5 6NT, UK}
\altaffiltext{7}{Konkoly Observatory of the Hungarian Academy of Sciences, Budapest, Hungary}

\begin{abstract}
The helical kink instability \q{of a twisted magnetic flux tube} has been suggested as a trigger mechanism for solar filament eruptions and coronal mass ejections (CMEs). 
\q{In order to investigate if estimations of the pre-eruptive twist can be obtained from observations of writhe in such events, we} 
quantitatively analyze the conversion of twist into writhe in the course of the instability, using numerical simulations. We consider the line tied, cylindrically symmetric Gold--Hoyle flux rope model and measure the writhe using the formulae by Berger and Prior which express the quantity as a single integral in space. We find that the amount of twist converted into writhe does not simply scale with the initial flux rope twist, but depends mainly on the growth rates of the instability eigenmodes of higher longitudinal order than the basic mode. 
\q{The saturation levels of the writhe, as well as the shapes of the kinked flux ropes, are very similar for considerable ranges of initial flux rope twists, which essentially precludes estimations of pre-eruptive twist from measurements of writhe. However, our simulations suggest an upper twist limit of $\sim 6\pi$ for the majority of filaments prior to their eruption.} 
\end{abstract}

\keywords{Magnetohydrodynamics (MHD); Sun: corona; Sun: filaments}

%%%%%%%%%%%%%%%%%%%%%%%%%%%%%%%%%%%%%%%%%%%%%%%%%%%%%%%%
\section{Introduction}
\label{s:int}
%%%%%%%%%%%%%%%%%%%%%%%%%%%%%%%%%%%%%%%%%%%%%%%%%%%%%%%%
The $m=1$ kink mode or helical kink instability (hereafter KI) is a current-driven, ideal magnetohydrodynamic (MHD) instability. It occurs in a magnetic flux rope if the winding of the field lines about the rope axis (the twist) exceeds a critical value \citep[e.g.,][]{shafranov57,kruskal58,freidberg82,priest82}. The instability lowers the magnetic energy of the flux rope by reducing the bending of field lines, which leads to a characteristic helical deformation (writhe) of the rope axis. Such writhing is often observed in erupting filaments or prominences in the solar corona (Figure\,\ref{f:filaments}), which has led to the suggestion that the KI can trigger filament eruptions and CMEs \citep[e.g.,][]{sakurai76,sturrock01,torok05,fan05}.

The KI has been studied extensively for laboratory plasmas \citep[see, e.g.,][and references therein]{bateman78,goedbloed10}. In applications relevant to the low-$\beta$ solar corona, typically force-free, cylindrically symmetric flux rope configurations of finite length are considered. The anchoring of coronal loops and prominences in the solar surface is modeled by imposing line tied boundary conditions at the flux rope ends. Properties of the KI such as the instability threshold and growth rate, as well as the formation of current sheets, have been investigated for various radial twist profiles in both straight and arched flux rope geometries \citep[e.g.,][]{hood81,mikic90,baty96,gerrard01,torok04}. MHD simulations of kink-unstable flux ropes have been employed to model coronal loop heating and 
\q{bright-point} 
emission \citep{galsgaard97,haynes07}, soft X-ray sigmoids \citep{kliem04}, energy release in compact flares \citep{gerrard03}, microwave sources in eruptive flares \citep{kliem10}, and rise profiles, rotation, and writhing of erupting filaments and CMEs \citep{torok05,williams05,fan05,kliem12}. In spite of this large body of work, the amount and evolution of the writhing in kink-unstable flux ropes was quantified only very rarely \citep{linton98,torok10}. Systematic investigations of the dependence of the writhe on parameters such as the initial flux rope twist or geometry have not yet been undertaken. 

The quantity \emph{writhe} measures the net 
\q{self-coiling} 
of a space curve. It is related to the total torsion along the curve: the sum of writhe and total torsion remains constant under deformations, unless the curve develops an inflexion point, where curvature vanishes \citep{moffatt92}. Twist and writhe of a thin flux rope are related to its magnetic helicity via $H=F^2(T+W)$, where $F$ is the axial magnetic flux, $T$ is the number of field line turns, and $W$ is the writhe of the rope axis \citep{calugareanu59,berger06}. The writhe for flux ropes with footpoints on a boundary (such as the photosphere) can be defined by the same formula, using relative helicity for $H$ \citep{berger84}. $W$ depends only on the shape of the axis of the rope; while $T$ measures the net twist of the field lines in the rope about the axis. Since magnetic helicity is conserved in ideal MHD, the KI converts twist into an equal amount of writhe. Here we quantify this process for the first time systematically for a range of initial flux rope twists, using MHD simulations. 
 
For our study we consider the straight, uniformly twisted, force-free flux rope equilibrium by \citet[][hereafter GH]{Gold60}, line tied at both ends. In the absence of knowledge about typical twist profiles in coronal flux ropes and due to its force freeness, the equilibrium serves as a convenient reference model. Mechanisms other than the KI that may cause writhing (see \citealt{kliem12} for a detailed discussion) are excluded. Furthermore, the KI of the GH model does not lead to the formation of a helical current sheet, which triggers reconnection in the nonlinear development of other flux rope equilibria \citep[e.g.,][]{baty96,gerrard03}. Therefore, the evolution of the axis deformation can be followed well into the saturation phase of the writhe, which makes this equilibrium particularly suited for our purpose. We measure the axis writhe using the formulae by \cite{berger06}, which express the quantity as a single integral in space, facilitating its calculation.

Our motivation for this study is derived from the interest in obtaining estimates of the twist in pre-eruptive solar configurations from the amount of writhing observed during an eruption. At present, the twist cannot be obtained directly, since the magnetic field cannot be measured in the coronal volume and since extrapolations from photospheric vector magnetograms are not yet sufficiently reliable in practice, especially for volumes containing a filament \citep{mcclymont97,schrijver08a}. Twist estimations based on the observations of pre-eruptive coronal configurations are hampered with substantial uncertainties (see Section\,\ref{s:dis}). The writhe of erupting filaments, on the other hand, can be obtained with a reasonable accuracy if the filament displays a coherent shape (Figure\,\ref{f:filaments}) and if observations from more than one viewing angle are available \citep[for example from the {\it STEREO} mission;][]{kaiser08} or if the eruption is directed toward the observer \citep{torok10}. Although twist estimates from the writhe can only be obtained in retrospect, they may facilitate systematic studies of this possibly critical parameter for CME initiation and may be useful for comparison with other means of estimation.

%%%%%%%%%%%%%%%%%%%%%%%%%%%%%%%%%%%%%%%%%%%%%%%%%%%%%%%
\section{Flux Rope Model and Numerical Setup}
\label{s:num}
%%%%%%%%%%%%%%%%%%%%%%%%%%%%%%%%%%%%%%%%%%%%%%%%%%%%%%%
The GH model used in this study is given by
\begin{equation}
 B_\theta = \frac{B_0 br}{1+b^2r^2},\,\,\,\,\,\, 
 B_z = \frac{B_0}{1+b^2r^2}
\end{equation}
and represents a uniformly twisted, force-free flux rope of infinite radial extent. The constant $b$ is related to the axial length of one field line turn, $\Lambda$, often referred to as the pitch, by $b=2\pi/\Lambda$ and, at the same time, represents the inverse scale length of the radial field profile. Using the customary form of the expression for the twist angle, 
\begin{equation}\label{eq:phi}
 \Phi(r)=\frac{L\,B_{\theta}(r)}{r\,B_z(r)},
\end{equation}
a GH rope of length $L$ has a twist of $\Phi=bL$. This is related to the number of field line turns $T=L/\Lambda$ by $\Phi=2\pi T$. In our calculations we fix the scale $b=\pi$, so that the radial field profiles are identical in all simulations, and we vary the initial twist by varying the flux rope length.

%_________________________________________________________________________
\begin{table}[b]
\centering
\begin{tabular}{|c|r|r|c|}
\hline    
$L_y$ & $\Phi_0$    & $h_0$ & $W$  \\
\hline
 1.48 & $ 3.0\,\pi$ &   4   & 0.96 \\
 2.24 & $ 4.5\,\pi$ &   4   & 0.97 \\
 3.00 & $ 6.0\,\pi$ &   4   & 1.01 \\
 3.76 & $ 7.5\,\pi$ &   6   & 1.04 \\
 4.48 & $ 9.0\,\pi$ &   8   & 1.79 \\
 5.28 & $10.6\,\pi$ &  10   & 1.76 \\
\hline   
\end{tabular}
\caption
{
Parameters of the simulations. 
$2 L_y$--length of the flux rope;
$\Phi_0=2\pi L_y$--initial twist; 
$h_0$--initial $z$ position of the rope axis;
$W$--writhe (peak writhe for $\Phi_0\ge9\pi$ and final writhe for $\Phi_0\le7.5\pi$).
}
\label{tbl:gh_runs}
\end{table}
%________________________________________________________________________

The numerical set--up for the simulations is the same as in our previous studies of the kink and torus instabilities \citep[e.g.,][]{torok05,torok07}; for a detailed description see \cite{torok03}. The compressible ideal MHD equations are integrated using the simplifying assumptions of vanishing pressure and gravity on a discretized Cartesian box $[-L_x,L_x]\times[-L_y,L_y]\times[0,L_z]$ with uniform spacing $\Delta=0.04$ and $L_x=8$, $L_z=16$. In order to take advantage of the $z$-axis line symmetry inherent in the configurations considered, we orient the flux rope parallel to the $y$ axis, so that the integration need be carried out only in the ``half box'' $\{y>0\}$. The flux rope length equals the full box length $2 L_y$. Except for $\{y=0\}$, where mirroring according to the $z$-axis line symmetry is applied, the MHD variables are held fixed at their initial values at all boundaries (and for consistency the velocity is kept at zero also one grid layer inside the boundaries). This models line-tying at the ends of the rope, $|y|=L_y$.

We vary the initial twist, $\Phi_0=2\pi L_y$, in the range (3--10.6)$\pi$ by varying $L_y$ in our series of runs. Since the rope expands strongly and in different ways in the different runs (Figure~\ref{fig:gh_rope}), we minimize the influence of the top and bottom boundaries by positioning the rope axis at appropriate initial heights $z=h_0$. See Table~\ref{tbl:gh_runs} for the values of $L_y$, $\Phi_0$, and $h_0$ used in this investigation.

The initial density distribution is specified to be 
\q{$\rho_0=B_0^{3/2}$}, 
such that the Alfv\'en velocity decreases slowly with distance from the flux rope axis (which corresponds to the conditions in the solar corona). The MHD variables are normalized by quantities derived from a characteristic length of the initial equilibria, chosen to be $\Lambda/2$, and the initial magnetic field strength and Alfv\'en velocity at the flux rope axis. All runs start with the fluid at rest. A small initial velocity  perturbation localized at the flux rope center is imposed in all runs \citep[analogous to][]{torok04}.

%%%%%%%%%%%%%%%%%%%%%%%%%%%%%%%%%%%%%%%%%%%%%%%%%%%%%%%
\section{Results}
\label{s:res}
%%%%%%%%%%%%%%%%%%%%%%%%%%%%%%%%%%%%%%%%%%%%%%%%%%%%%%%
Since the chosen initial twists in the series all exceed the KI threshold for the line tied GH equilibrium of $\Phi_\mathrm{cr}\approx2.5\pi$ \citep{hood81}, all configurations are unstable. The helical nature of the growing perturbation is clearly visible in the linear phase (the early phase of the instability during which the exponentially growing amplitude of the axis displacement remains small; top two rows in Figure~\ref{fig:gh_rope}). In the nonlinear phase (when the amplitude 
\q{of the axis displacement} 
becomes large) the flux rope starts to expand strongly by the action of the hoop force, which comes into play as soon as the flux rope develops some overall curvature between its line tied ends (bottom two rows in Figure~\ref{fig:gh_rope} and Section~\ref{s:dis}).

%------------------------------------------------------------------------------------ 
\subsection{Flux rope axis evolution}
%------------------------------------------------------------------------------------
We measure the growth of writhe at the axis of the flux rope using Equations\,(4)--(6) in \cite{torok10}; see also \cite{berger06}. Note that flux surfaces away from the axis undergo a smaller deformation, with less twist converted into writhe, but Figure~\ref{fig:gh_rope} indicates that a substantial cross section of the GH rope attains similar writhe. For our strongly twisted cases, the
\q{measurements are reliable} 
only until the perturbed flux surfaces start to approach the boundaries of the box.

Figure~\ref{fig:gh_writhe}\q{(a)} 
shows the development of writhe by the KI. The writhe grows exponentially in the linear phase and then reaches saturation in the nonlinear phase of the instability. It can be seen that the saturation level of the writhe does not scale linearly with the initial twist, which is different from what one might intuitively expect.

The flux ropes with the smallest twist, $\Phi_0=3\pi$ and $4.5\pi$, exhibit a very similar behavior, except for a significantly faster initial evolution of the run with $\Phi_0=4.5\pi$. The writhe saturates at $W \approx 0.95$ in both runs, corresponding to a converted twist of $\Phi \approx 1.9\pi$ in the vicinity of the flux rope axis. The morphological evolution and the resulting axis shapes are very similar too (Figure~\ref{fig:gh_rope}). In both cases, the axis deforms into a one-turn helix. 
\q{Figure~\ref{fig:gh_writhe}(b) shows that the evolution of writhe coincides well with the release of magnetic energy by the KI and the displacement of the flux rope axis.}

A somewhat different evolution takes place for $\Phi_0=7.5\pi$: the writhe first reaches a maximum after the initial exponential growth phase, then decreases 
\q{by $\approx 20$ percent,}
but subsequently starts to increase again slowly, reaching $W \approx 1.05$ at the end of the simulation. 
\q{Figure~\ref{fig:gh_writhe}(b) shows that the flux rope continues to rise (at a slower rate) after the writhe has reached its first maximum, accompanied by ongoing magnetic energy release. The energy saturates when the writhe reaches its temporary minimum. Some further release occurs later on in the evolution; this appears to be related to reconnection that occurs at outer flux surfaces when those approach the boundary of the simulation box.} 
The morphological evolution is somewhat different from the smaller-twist cases: the axis shape obtained in the nonlinear phase of the instability is still dominated by a one-turn helix, but it becomes internally helically deformed (see Figure~\ref{fig:gh_rope}). The decrease of the writhe appears to be related to the reversal of the orientation of the flux rope legs in the vicinity of the $\pm L_y$ boundaries 
(\q{see} 
the second and fourth panel for $\Phi_0=7.5\,\pi$ in Figure~\ref{fig:gh_rope}), which occurs at the same time as the writhe decrease. This is a consequence of the line-tying that would be absent in infinitely extended flux ropes, and it implies a temporary increase of twist in the vicinity of the flux rope axis. The subsequent increase of the writhe is most likely associated with the development of the internal axis deformation,
\q{but may be related to some degree also to the reconnection mentioned above.}

The run with $\Phi_0=6.0\,\pi$ is an intermediate case: both the temporary decrease of the writhe and the internal helical axis deformation are present but rather weak, and the writhe at the end of the simulation ($W \approx 1.0$) lies in between the corresponding values for the smaller-twist runs and the run with $\Phi_0=7.5\pi$.

The cases with the largest twists ($\Phi_0=9\pi$ and $10.6\pi$) show a very different behavior. After the fast initial rise, the writhe continues to grow at a much smaller rate until it reaches a maximum of $W \approx 1.75-1.8$, after which it slowly decreases (the decrease is not visible for $\Phi_0=9\pi$ in the figure, since the evolution during the nonlinear phase is significantly slower than for $\Phi_0=10.6\pi$). The flux rope axis now develops a helix with about two turns (see Figure~\ref{fig:gh_rope} for $\Phi_0=9\,\pi$). We attribute the slow increase and decrease of the writhe to the complex morphological evolution of the flux rope for large twists: the development of two expanding helices within the finite domain forces approaching flux rope sections to give way to one another -- an effect that is much less pronounced in cases where only one helix develops. The decrease may be also a consequence of the strongly expanding helices approaching the boundaries of the simulation box.

%------------------------------------------------------------------------------------ 
\subsection{What determines the amount of twist converted into writhe?}
%------------------------------------------------------------------------------------
From 
Figure~\ref{fig:gh_writhe}\q{(a)} 
it is obvious that the conversion of twist into writhe depends on the initial twist in a non-trivial manner. Figures~\ref{fig:gh_rope} and \ref{fig:gh_writhe} indicate that it is related to the number of helical turns the rope axis develops in the course of the instability. This number depends on the wavelength of the most unstable mode and on the range of unstable modes permitted by the finite length of the rope. In the linear phase of the instability, the helical eigenmode with the highest growth rate dominates the way the rope starts to deform. The finite length of the line tied rope modifies this picture in the nonlinear phase.

The growth rate as a function of axial wavelength for the GH equilibrium is shown in Figure~\ref{fig:gh_growth} \citep[from][]{linton98}. This plot is for the case of infinite axial but finite radial extent of the rope, $R=3\pi/2b$. \cite{linton98} find the peak growth rate and its location to be unchanged for larger $R$ (our $x$-$z$ box sizes correspond to $R\sim8\pi/b$). The growth of the helical kink mode peaks at the wavelength $\lambda=1.85 \Lambda$, meaning that the KI grows fastest at a writhe wavelength of about twice the twist pitch. For our choice $b=\pi$ we have $2\Lambda=4$. Therefore, for a double helix to dominate in the linear phase, the required box length is $2L_y>8$, equivalent to a required twist $\Phi=2\pi L_y>8\pi$.

Although this corresponds nicely to the jump of the final writhe in Figure~\ref{fig:gh_rope} between twists $\Phi_0=7.5\pi$ and $9\pi$, it does not actually explain the occurrence of a jump. From the stability analysis we know that the most rapidly growing mode at $\lambda\approx2\Lambda$ is permitted to occur as soon as the box length satisfies $2L_y>2\Lambda=4$. Therefore, for $2L_y>2\Lambda=4$ non-integer values for the number of turns of $\approx L_y/\Lambda$ can occur at the peak growth rate in the linear phase. This agrees with the simulation results shown in the two upper rows of Figure~\ref{fig:gh_rope}. The dominant mode in this phase exhibits a little more than one turn for $\Phi_0=4.5\pi$, nearly two turns for $\Phi_0=7.5\pi$ and a little more than two turns for $\Phi_0=9\pi$.

The jump in the writhe can only be understood from the nonlinear evolution of the instability. This shows a clear tendency to develop an integer number of turns. As long as two axial wavelengths don't fit into the box, the mode with a single turn dominates in this phase. Contributions of the linearly most strongly growing mode, which then has a shorter wavelength, are clearly present (most obviously for $\Phi_0=7.5\pi$), but no longer dominate. We attribute this result to the action of the hoop force for kinking flux ropes of finite length. The line-tying leads to an axial dependence of the displacement which is absent for infinitely extended ropes. For $\Phi_0\le7.5\pi$ the displacement is largest in the mid-plane $\{y=0\}$ of our symmetric simulations and tapers off toward the line tied ends at $y=\pm L_y$ (see the two upper rows in Figure~\ref{fig:gh_rope}). An overall net bending of the rope in the direction of the displacement in the mid-plane (which points along the $z$ axis in all our simulations) results. This introduces a Lorentz self-force in the rope, known as hoop force, pointing in the direction of the bending \citep{bateman78}. The hoop force amplifies the perturbation of the rope axis in the mid-plane above its purely KI-driven displacement, thus creating a one-turn helix for $\Phi_0\le7.5\pi$. Such amplification occurs at a pair of symmetrically located displacements for $\Phi_0\ge9\pi$, creating a helix with two turns. We discuss the implications of this result for filament eruptions in the following section.

%%%%%%%%%%%%%%%%%%%%%%%%%%%%%%%%%%%%%%%%%%%%%%%%%%%%%%% 
\section{Discussion}
\label{s:dis}
%%%%%%%%%%%%%%%%%%%%%%%%%%%%%%%%%%%%%%%%%%%%%%%%%%%%%%%

We studied the conversion of twist into writhe in a simulation series of the KI in the GH model, considering initial flux rope twists in the range $3.0\,\pi \le \Phi_0 \le 10.6\,\pi$. We found in all cases a saturation of writhe in the nonlinear phase of the instability, after an initial exponential increase during the linear phase. However, the final writhe does not scale simply with the initial flux rope twist. Rather, the amount of twist converted into writhe seems to be determined predominantly by the number of helical turns the flux rope axis develops in the nonlinear phase. 

For $3.0\,\pi \lesssim \Phi_0 \lesssim 7.5\,\pi$, the rope axis develops a one-turn helix. For twists close to the upper end of this range, internal helical deformations of the one-turn helix develop, due to helical eigenmodes with wavelengths $\lambda < 2 L_y.$ However, the axis shape remains to be dominated by one turn in the nonlinear phase of the instability. The resulting writhe is close to unity for all cases, corresponding to a converted twist of $\sim 2\,\pi$ in the vicinity of the flux rope axis. If the twist is increased beyond this range, the rope axis develops more than one helical turn, and considerably more twist is converted into writhe ($\sim 3.5\,\pi$ in our simulations with $\Phi_0=9.0\,\pi$ and $\Phi_0=10.6\,\pi$). We attributed the relatively similar writhe values obtained in each respective range, as well as the pronounced increase of the writhe between them, to the action of the hoop force on line-tied, kink-unstable flux ropes of finite length. 

The basically discontinuous dependence of the final writhe upon the initial twist displayed in Figure~\ref{fig:gh_writhe} essentially precludes a reasonable estimation of the initial twist from observations of the writhe in solar filament eruptions and CMEs. The saturation levels of the writhe are \q{very} similar for initial twists up to $\Phi_0\approx8\pi$, requiring an accuracy of writhe determination for such an estimate that cannot be reached in solar observations. Moreover, the final writhe, as any other property of the KI, depends on the radial twist profile of the initial equilibrium. Therefore, a precise knowledge of this profile, combined with a parametric simulation study like the one in Figures~\ref{fig:gh_rope} and \ref{fig:gh_writhe} for a range of different profiles, would be required to permit a reliable estimate of twist. Further effects of importance for the final writhe enter when arched flux rope equilibria are considered \citep[see][]{torok10}, rendering a twist estimation from writhe observations even more difficult.

In order to compare our results with the KI in force-free equilibria with non-uniform radial twist
\q{profile},
we performed simulation series similar to the one presented here for the straight flux rope model termed ``Equilibrium 2'' in \cite{gerrard01} and the arched flux rope model by \cite{titov99}. Unfortunately, in all runs the flux rope axis was destroyed by reconnection at current sheets before the writhe would clearly saturate (see \citealt{amari99a}, \citealt{haynes08}, and \citealt{valori10} for examples of such reconnection), so that these simulations cannot be used for the purpose of this study.
   
However, our writhe measurements for the KI in the GH equilibrium provide at least a rough upper limit for the initial twist of erupting filaments. It is observed that kinking filaments typically do not display more than one helical turn and hardly any significant internal helical deformation of their axis. Combined with our simulations, this suggests that the initial twist typically does not exceed values $\Phi_0\sim6\pi$. This is supported by simulations of the KI in the Titov-D\'emoulin model, which show strong internal helical deformations for twists above this value (see, e.g., Figure~1 in \citealt{kliem10} and Figure~12 in \citealt{torok10}).
 
\begin{figure*}[t]
\centering
\includegraphics[width=1.\linewidth]{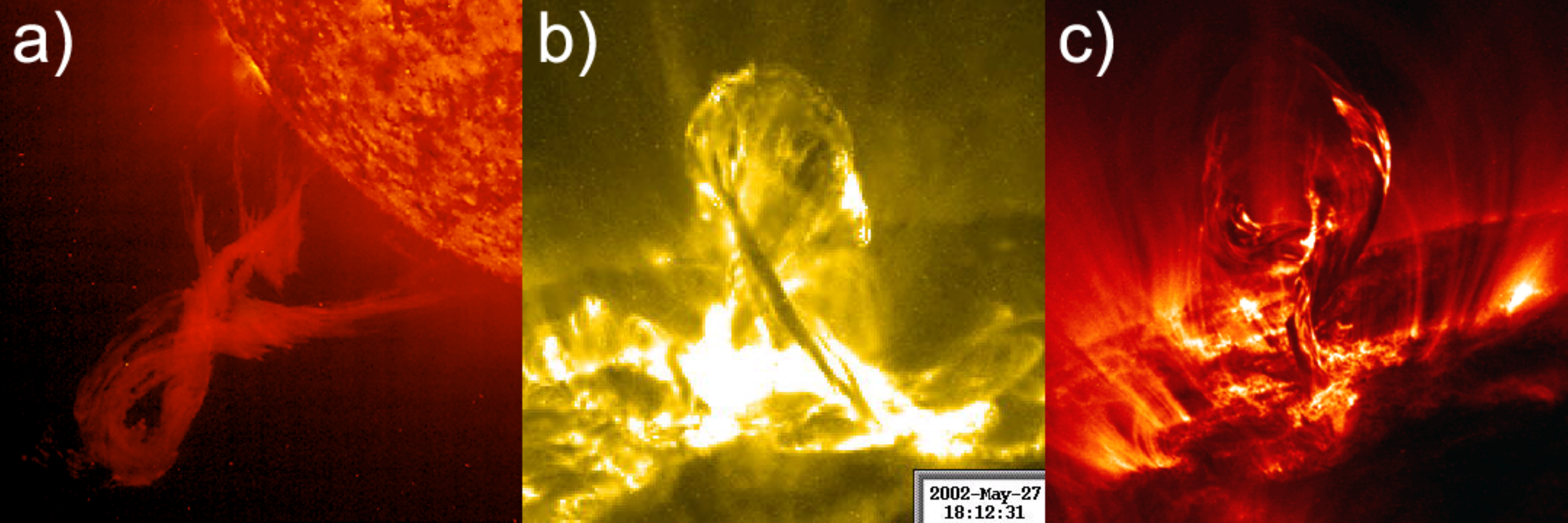}
\caption
{
Erupting and writhing solar filaments observed in extreme ultraviolet (EUV) 
wavelengths. 
{\bf a):} A full eruption (evolving into a CME) on 18 January 2000, observed 
          in 304 \AA\, by the EIT telescope onboard the SOHO spacecraft.  
{\bf b):} A confined eruption (trapped in the low corona) on 27 May 2002, 
          observed in 195 \AA\, by the TRACE satellite.
{\bf c):} An eruption, which most likely 
\q{remained}
          confined, on 19 July 2000, observed in 171 \AA\, by TRACE.
}
\label{f:filaments}
\end{figure*}

\begin{figure*}[t]
\centering
\includegraphics[width=0.75\linewidth]{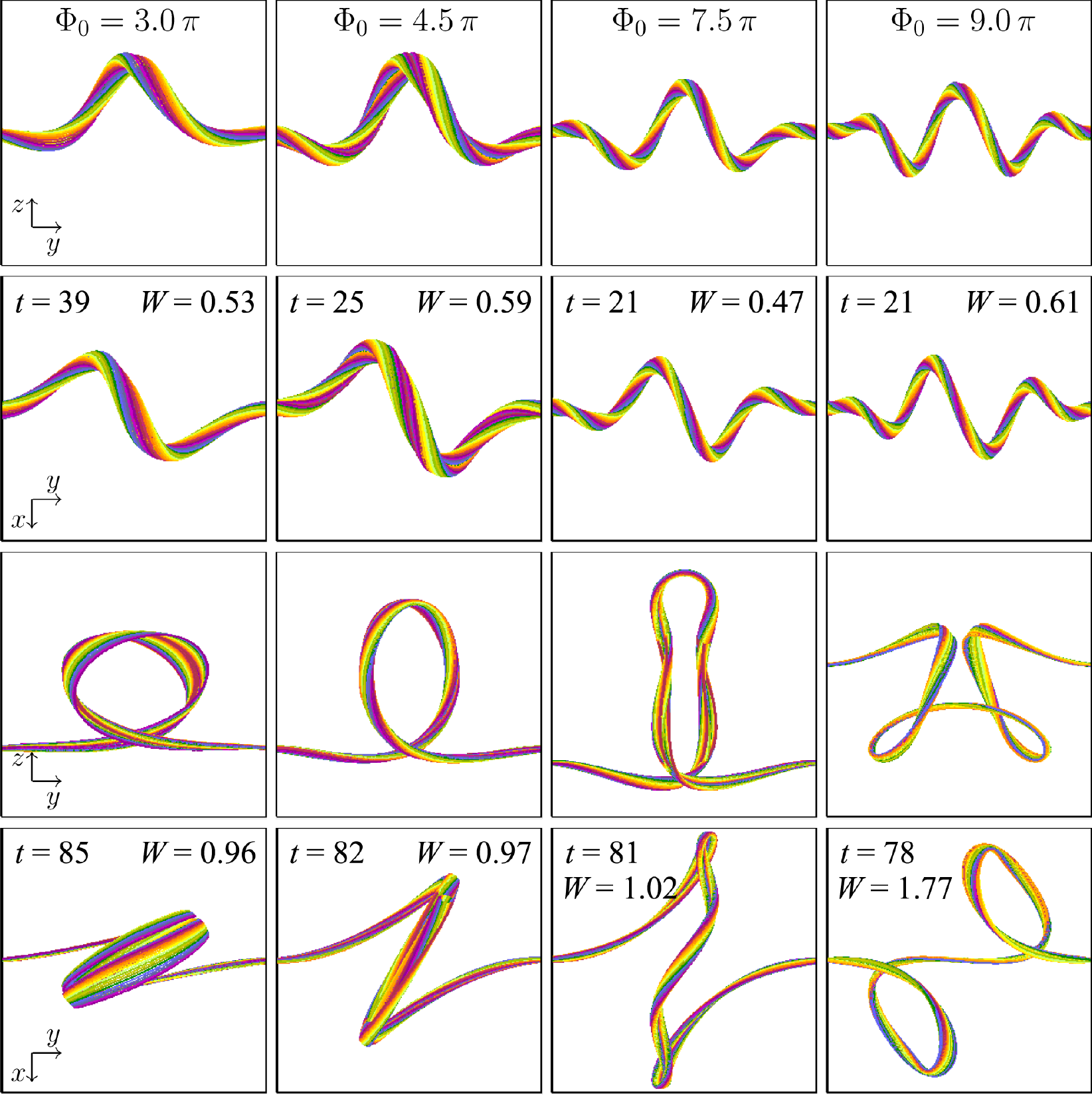}
\caption
{ 
Helical deformation of the kink-unstable Gold-Hoyle flux rope. Magnetic 
field lines start at circles of radius $R$ centered at the axis endpoints, 
$(0,\pm L_y,h_0)$. The full box range in $y$ direction, $-L_y \le y \le L_y$, 
is shown in all panels. Endpoint locations are the same for each column, 
while the $x$ and $z$ ranges are changed to account for the expansion of
the flux rope. The two upper rows show flux surfaces with $R=0.1$ during 
or shortly after the linear phase of the instability in a side view and a
top view, respectively. The range $h_0-2 \le z \le h_0+2$ is displayed in
the side views and $-L_y \le x \le L_y$ in the top views. The two lower
rows show flux surfaces with $R=0.05$ at the end, or toward the end, of 
the simulations in the same views, using the vertical ranges $[2,10]$,
$[2,10]$, $[4,14]$, and $[0,14]$ (from left to right) in the side views
and $-L_y\le x \le L_y$ in the top views. Simulation times and the value 
of writhe are shown in the top view panels for all cases.
}
\label{fig:gh_rope}
\end{figure*}

\begin{figure*}[t]
\centering
\includegraphics[width=1.\linewidth]{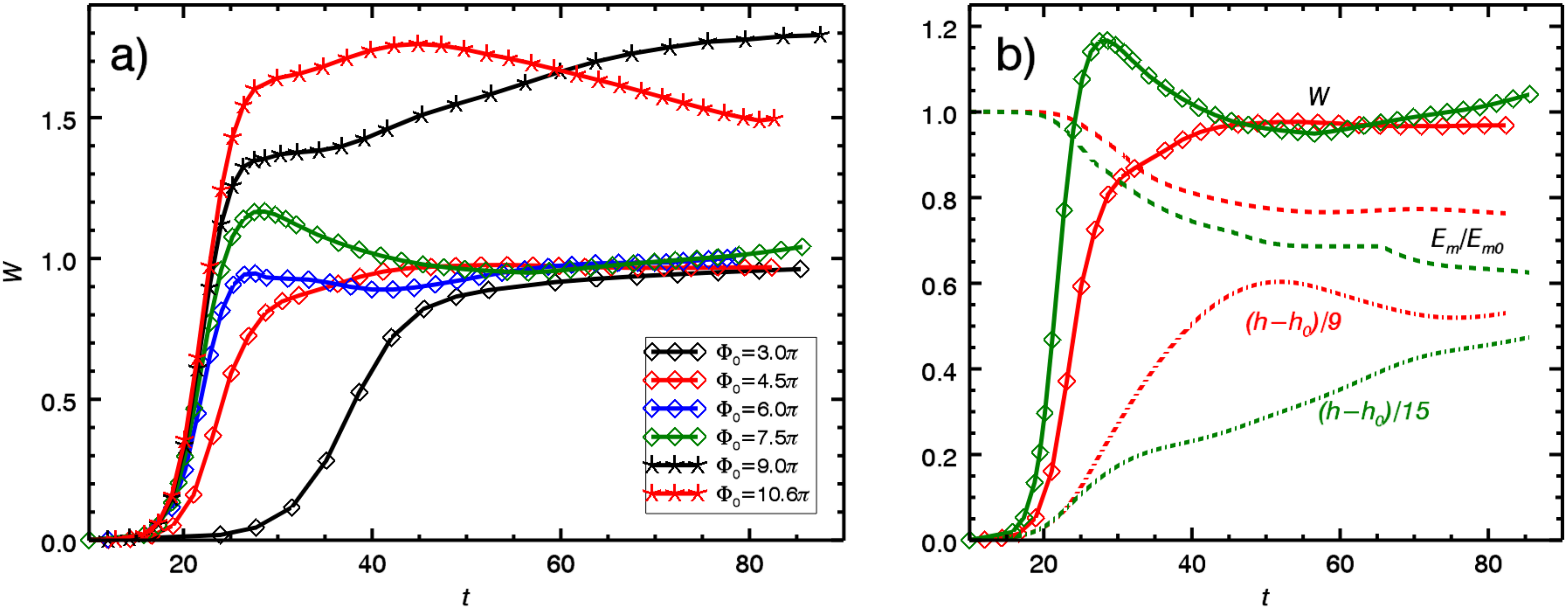}
\caption{
{\bf (a):} Development of writhe by the KI of the GH flux rope for different 
           values of the initial twist.
{\bf (b):} Writhe 
\q{(diamonds),} 
           vertical displacement of the flux rope axis from its initial position 
           on the $z$ axis 
\q{(dash-dotted lines}; 
           scaled to fit into the plot), and total magnetic energy
\q{(dashed lines}; 
           normalized to initial value) for $\Phi_0=4.5\,\pi$ 
\q{(red) and $\Phi_0=7.5\,\pi$ (green)}.
}
\label{fig:gh_writhe}
\end{figure*}

\begin{figure}[t]
\centering
\includegraphics[width=1.0\linewidth]{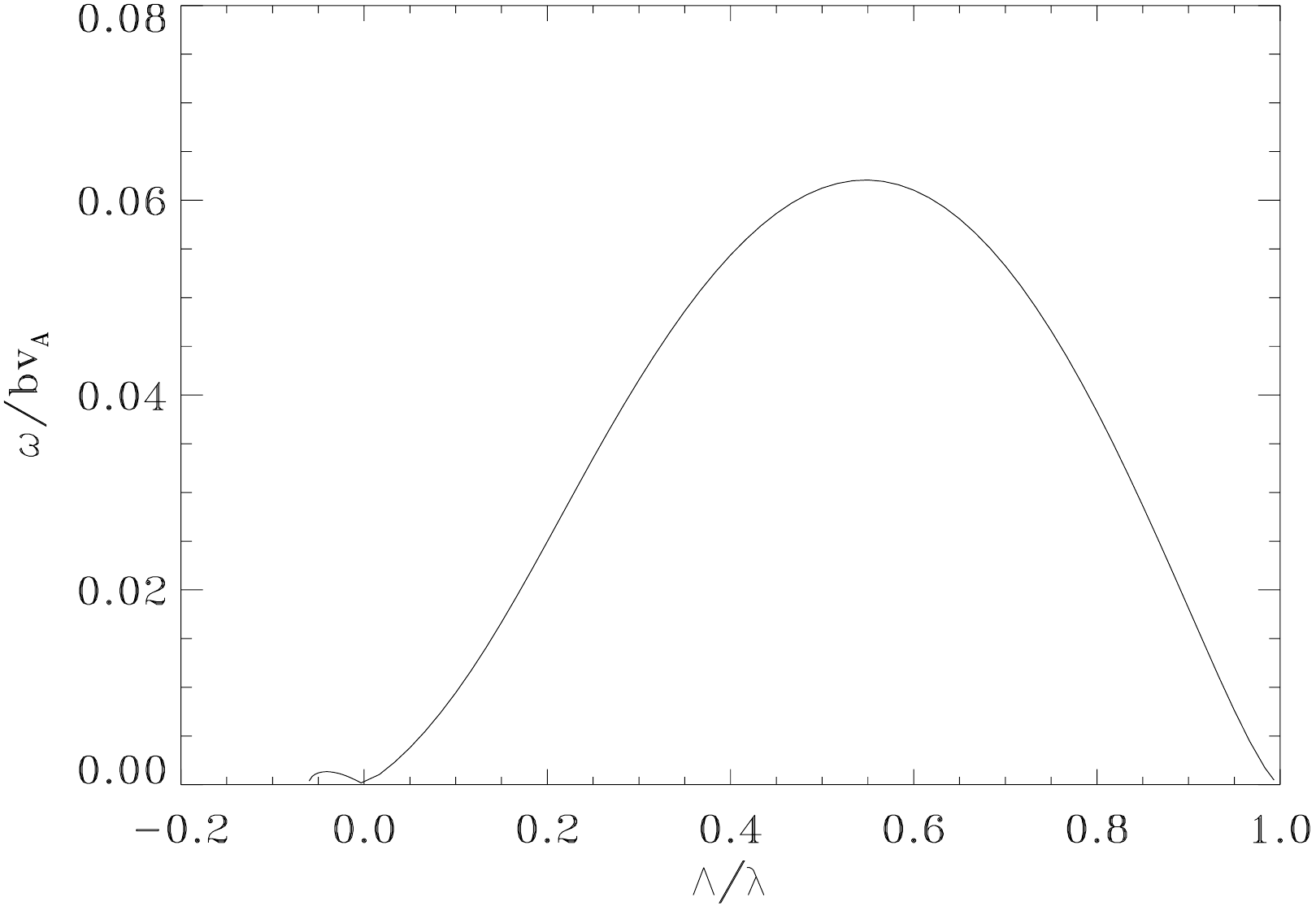}
\caption
{
Helical kink instability growth rates $\omega$ vs.\ axial wavelength $\Lambda/\lambda$ 
for the Gold-Hoyle equilibrium of infinite axial but finite radial extent of $R=3\pi/2b$ 
(from Linton {\it et al.}, 1998;
\q{\copyright\,AAS. Reproduced with permission).}
$b=B_\theta/(rB_z)$ is the twist per unit length and $v_A$ is the initial Alfv\'en 
velocity at the flux rope axis.
}
\label{fig:gh_growth}
\end{figure}

Occasionally, however, the Sun seems to succeed in building up higher twists. Several examples can be found in \cite{vrsnak91}, whose estimates of the end-to-end twist fall in the range $(3\mbox{--}15)\,\pi$ for a sample of prominences close to the time of eruption. A particularly clear indication of very high twist (of $\sim10\pi$) was obtained by \cite{romano03} for a filament eruption on 19 July 2000 (Figure\,\ref{f:filaments}c). These estimations are based on measurements of the pitch angle of selected helical prominence threads, which are then converted into twist assuming a uniform radial twist profile both along and across the axis of the underlying flux rope. The latter assumption may be a severe oversimplification, since force-free flux ropes embedded in potential field must generally have a nonuniform radial twist profile in order to match the field at the surface of the rope \citep[see, e.g., Figure~2 in][]{torok04}. Still, the simulations presented here support the existence of such a high twist at least for the case shown in Figure\,\ref{f:filaments}c, based on the strong bending in the lower part of the filament legs \citep[see also Figure 12b in][where a strongly nonuniform radial twist profile was used]{torok10}. A further observed case of very high twist may have been an apparently three-fold helix described in \cite{gary04}.

While these estimations remain uncertain to a considerable degree, we can ask how such large twists, if present, may be produced in the solar corona. It is widely believed that twist is accumulated prior to an eruption by flux emergence \citep[e.g.,][]{leka96}, photospheric vortex flows \citep[e.g.,][]{romano05}, or the slow transformation of a sheared magnetic arcade into a flux rope \citep[e.g.,][]{moore01,aulanier10}. It has been argued that flux ropes that form by one or more of these mechanisms will become kink-unstable long before the large twists mentioned above can be reached. While this is likely true for the majority of cases, several scenarios for the build-up of large twists appear to plausible. First, the KI threshold can vary in a wide range as a function of the thickness of the rope \citep[e.g.,][]{hood79,baty01,torok04}, so sufficiently thin flux ropes may be able to harbor large twist in a stable state. Second, sufficiently flat highly twisted flux ropes may be stabilized by strong ambient shear fields \citep[][]{torok10}, or by gravity if they contain sufficient filament material. Third, significant twist may be added by reconnection to the rising flux in the course of an eruption \citep[e.g.,][]{qiu07}. Finally, flux ropes may reconnect and merge prior to an eruption, thereby adding up their respective twists \citep[e.g.,][]{pevtsov96,canfield98,schmieder04,vanballegooijen04}.

\acknowledgments
We thank the anonymous referees for very helpful suggestions and Z. Miki\'c for providing a routine that was helpful for the writhe calculations. The contribution of T.T. was supported by NASA's HTP, LWS, and SR\&T programs and by the NSF. M.G.L. received support from NASA/LWS and the ONR 6.1 programs. B.K. was supported by the DFG. The research leading to these results has received funding from the European Commission's Seventh Framework Programme under the grant agreement No. 284461 (eHEROES project). L.vDG.'s work work was supported by the STFC Consolidated Grant
ST/H00260X/1 and the Hungarian Research grant OTKA K-081421.

% format of references provided by the journal (.bst)
%\bibliographystyle{apj}
% name your Bibtex file containing your references (.bib)
%\bibliography{torok}  

\begin{thebibliography}{52}
\expandafter\ifx\csname natexlab\endcsname\relax\def\natexlab#1{#1}\fi

\bibitem[{{Amari} \& {Luciani}(1999)}]{amari99a}
{Amari}, T., \& {Luciani}, J.~F. 1999, \apjl, 515, L81

\bibitem[{{Aulanier} {et~al.}(2010){Aulanier}, {T{\"o}r{\"o}k}, {D{\'e}moulin},
  \& {DeLuca}}]{aulanier10}
{Aulanier}, G., {T{\"o}r{\"o}k}, T., {D{\'e}moulin}, P., \& {DeLuca}, E.~E.
  2010, \apj, 708, 314

\bibitem[{{Bateman}(1978)}]{bateman78}
{Bateman}, G. 1978, {MHD instabilities} (Cambridge, Mass., MIT Press)

\bibitem[{{Baty}(2001)}]{baty01}
{Baty}, H. 2001, \aap, 367, 321

\bibitem[{{Baty} \& {Heyvaerts}(1996)}]{baty96}
{Baty}, H., \& {Heyvaerts}, J. 1996, \aap, 308, 935

\bibitem[{{Berger} \& {Field}(1984)}]{berger84}
{Berger}, M.~A., \& {Field}, G.~B. 1984, Journal of Fluid Mechanics, 147, 133

\bibitem[{{Berger} \& {Prior}(2006)}]{berger06}
{Berger}, M.~A., \& {Prior}, C. 2006, Journal of Physics A Mathematical
  General, 39, 8321

\bibitem[{{Canfield} \& {Reardon}(1998)}]{canfield98}
{Canfield}, R.~C., \& {Reardon}, K.~P. 1998, \solphys, 182, 145

\bibitem[{C\u{a}lug\u{a}reanu(1959)}]{calugareanu59}
C\u{a}lug\u{a}reanu. 1959, Czechoslovak Math. J., 11, 588{\~n}625

\bibitem[{{Fan}(2005)}]{fan05}
{Fan}, Y. 2005, \apj, 630, 543

\bibitem[{{Freidberg}(1982)}]{freidberg82}
{Freidberg}, J.~P. 1982, Reviews of Modern Physics, 54, 801

\bibitem[{{Galsgaard} \& {Nordlund}(1997)}]{galsgaard97}
{Galsgaard}, K., \& {Nordlund}, {\AA}. 1997, \jgr, 102, 219

\bibitem[{{Gary} \& {Moore}(2004)}]{gary04}
{Gary}, G.~A., \& {Moore}, R.~L. 2004, \apj, 611, 545

\bibitem[{{Gerrard} {et~al.}(2001){Gerrard}, {Arber}, {Hood}, \& {Van der
  Linden}}]{gerrard01}
{Gerrard}, C.~L., {Arber}, T.~D., {Hood}, A.~W., \& {Van der Linden}, R.~A.~M.
  2001, \aap, 373, 1089

\bibitem[{{Gerrard} \& {Hood}(2003)}]{gerrard03}
{Gerrard}, C.~L., \& {Hood}, A.~W. 2003, \solphys, 214, 151

\bibitem[{{Goedbloed} {et~al.}(2010){Goedbloed}, {Keppens}, \&
  {Poedts}}]{goedbloed10}
{Goedbloed}, J.~P., {Keppens}, R., \& {Poedts}, S. 2010, {Advanced
  Magnetohydrodynamics} (Cambridge University Press)

\bibitem[{{Gold} \& {Hoyle}(1960)}]{Gold60}
{Gold}, T., \& {Hoyle}, F. 1960, \mnras, 120, 89

\bibitem[{{Haynes} \& {Arber}(2007)}]{haynes07}
{Haynes}, M., \& {Arber}, T.~D. 2007, \aap, 467, 327

\bibitem[{{Haynes} {et~al.}(2008){Haynes}, {Arber}, \& {Verwichte}}]{haynes08}
{Haynes}, M., {Arber}, T.~D., \& {Verwichte}, E. 2008, \aap, 479, 235

\bibitem[{{Hood} \& {Priest}(1979)}]{hood79}
{Hood}, A.~W., \& {Priest}, E.~R. 1979, \solphys, 64, 303

\bibitem[{{Hood} \& {Priest}(1981)}]{hood81}
---. 1981, Geophysical and Astrophysical Fluid Dynamics, 17, 297

\bibitem[{{Kaiser} {et~al.}(2008){Kaiser}, {Kucera}, {Davila}, {St.~Cyr},
  {Guhathakurta}, \& {Christian}}]{kaiser08}
{Kaiser}, M.~L., {Kucera}, T.~A., {Davila}, J.~M., {St.~Cyr}, O.~C.,
  {Guhathakurta}, M., \& {Christian}, E. 2008, \ssr, 136, 5

\bibitem[{{Kliem} {et~al.}(2010){Kliem}, {Linton}, {T{\"o}r{\"o}k}, \&
  {Karlick{\'y}}}]{kliem10}
{Kliem}, B., {Linton}, M.~G., {T{\"o}r{\"o}k}, T., \& {Karlick{\'y}}, M. 2010,
  \solphys, 266, 91

\bibitem[{{Kliem} {et~al.}(2004){Kliem}, {Titov}, \& {T{\"o}r{\"o}k}}]{kliem04}
{Kliem}, B., {Titov}, V.~S., \& {T{\"o}r{\"o}k}, T. 2004, \aap, 413, L23

\bibitem[{{Kliem} {et~al.}(2012){Kliem}, {T{\"o}r{\"o}k}, \&
  {Thompson}}]{kliem12}
{Kliem}, B., {T{\"o}r{\"o}k}, T., \& {Thompson}, W.~T. 2012, \solphys, 281, 137

\bibitem[{{Kruskal} \& {Tuck}(1958)}]{kruskal58}
{Kruskal}, M., \& {Tuck}, J.~L. 1958, Royal Society of London Proceedings
  Series A, 245, 222

\bibitem[{{Leka} {et~al.}(1996){Leka}, {Canfield}, {McClymont}, \& {van
  Driel-Gesztelyi}}]{leka96}
{Leka}, K.~D., {Canfield}, R.~C., {McClymont}, A.~N., \& {van Driel-Gesztelyi},
  L. 1996, \apj, 462, 547

\bibitem[{{Linton} {et~al.}(1998){Linton}, {Dahlburg}, {Fisher}, \&
  {Longcope}}]{linton98}
{Linton}, M.~G., {Dahlburg}, R.~B., {Fisher}, G.~H., \& {Longcope}, D.~W. 1998,
  \apj, 507, 404

\bibitem[{{McClymont} {et~al.}(1997){McClymont}, {Jiao}, \& {Miki{\'c}}}]{mcclymont97}
{McClymont}, A.~N., {Jiao}, L., \& 
%
\q{{Miki{\'c}}}, 
%
Z. 1997, \solphys, 174, 191

\bibitem[{
%
{Miki{\'c}}
%
{et~al.}(1990){Miki{\'c}}, {Schnack}, \& {van Hoven}}]{mikic90}
%
\q{{Miki{\'c}}}, 
%
Z., {Schnack}, D.~D., \& {van Hoven}, G. 1990, \apj, 361, 690

\bibitem[{Moffatt \& Ricca(1992)}]{moffatt92}
Moffatt, H.~K., \& Ricca, R.~L. 1992, Proc. Roy. Soc. A, 439, 411

\bibitem[{{Moore} {et~al.}(2001){Moore}, {Sterling}, {Hudson}, \&
  {Lemen}}]{moore01}
{Moore}, R.~L., {Sterling}, A.~C., {Hudson}, H.~S., \& {Lemen}, J.~R. 2001,
  \apj, 552, 833

\bibitem[{{Pevtsov} {et~al.}(1996){Pevtsov}, {Canfield}, \&
  {Zirin}}]{pevtsov96}
{Pevtsov}, A.~A., {Canfield}, R.~C., \& {Zirin}, H. 1996, \apj, 473, 533

\bibitem[{{Priest}(1982)}]{priest82}
{Priest}, E.~R. 1982, {Solar magneto-hydrodynamics} (Dordrecht, Holland;
  Boston: D.~Reidel Pub.~Co.; Hingham)

\bibitem[{{Qiu} {et~al.}(2007){Qiu}, {Hu}, {Howard}, \& {Yurchyshyn}}]{qiu07}
{Qiu}, J., {Hu}, Q., {Howard}, T.~A., \& {Yurchyshyn}, V.~B. 2007, \apj, 659,
  758

\bibitem[{{Romano} {et~al.}(2003){Romano}, {Contarino}, \&
  {Zuccarello}}]{romano03}
{Romano}, P., {Contarino}, L., \& {Zuccarello}, F. 2003, \solphys, 214, 313

\bibitem[{{Romano} {et~al.}(2005){Romano}, {Contarino}, \&
  {Zuccarello}}]{romano05}
---. 2005, \aap, 433, 683

\bibitem[{{Sakurai}(1976)}]{sakurai76}
{Sakurai}, T. 1976, \pasj, 28, 177

\bibitem[{{Schmieder} {et~al.}(2004){Schmieder}, {Mein}, {Deng}, {Dumitrache},
  {Malherbe}, {Staiger}, \& {Deluca}}]{schmieder04}
{Schmieder}, B., {Mein}, N., {Deng}, Y., {Dumitrache}, C., {Malherbe}, J.,
  {Staiger}, J., \& {Deluca}, E.~E. 2004, \solphys, 223, 119

\bibitem[{{Schrijver} {et~al.}(2008){Schrijver}, {Elmore}, {Kliem},
  {T{\"o}r{\"o}k}, \& {Title}}]{schrijver08a}
{Schrijver}, C.~J., {Elmore}, C., {Kliem}, B., {T{\"o}r{\"o}k}, T., \& {Title},
  A.~M. 2008, \apj, 674, 586

\bibitem[{{Shafranov}(1957)}]{shafranov57}
{Shafranov}, V.~D. 1957, Journal of Nuclear Energy II, 5, 86

\bibitem[{{Sturrock} {et~al.}(2001){Sturrock}, {Weber}, {Wheatland}, \&
  {Wolfson}}]{sturrock01}
{Sturrock}, P.~A., {Weber}, M., {Wheatland}, M.~S., \& {Wolfson}, R. 2001,
  \apj, 548, 492

\bibitem[{{Titov} \& {D{\'e}moulin}(1999)}]{titov99}
{Titov}, V.~S., \& {D{\'e}moulin}, P. 1999, \aap, 351, 707

\bibitem[{{T{\"o}r{\"o}k} {et~al.}(2010){T{\"o}r{\"o}k}, {Berger}, \&
  {Kliem}}]{torok10}
{T{\"o}r{\"o}k}, T., {Berger}, M.~A., \& {Kliem}, B. 2010, \aap, 516, A49

\bibitem[{{T{\"o}r{\"o}k} \& {Kliem}(2003)}]{torok03}
{T{\"o}r{\"o}k}, T., \& {Kliem}, B. 2003, \aap, 406, 1043

\bibitem[{{T{\"o}r{\"o}k} \& {Kliem}(2005)}]{torok05}
---. 2005, \apjl, 630, L97

\bibitem[{{T{\"o}r{\"o}k} \& {Kliem}(2007)}]{torok07}
---. 2007, \an, 328, 743

\bibitem[{{T{\"o}r{\"o}k} {et~al.}(2004){T{\"o}r{\"o}k}, {Kliem}, \&
  {Titov}}]{torok04}
{T{\"o}r{\"o}k}, T., {Kliem}, B., \& {Titov}, V.~S. 2004, \aap, 413, L27

\bibitem[{{Valori} {et~al.}(2010){Valori}, {Kliem}, {T{\"o}r{\"o}k}, \&
  {Titov}}]{valori10}
{Valori}, G., {Kliem}, B., {T{\"o}r{\"o}k}, T., \& {Titov}, V.~S. 2010, \aap,
  519, A44+

\bibitem[{{van Ballegooijen}(2004)}]{vanballegooijen04}
{van Ballegooijen}, A.~A. 2004, \apj, 612, 519

\bibitem[{{Vr{\v s}nak} {et~al.}(1991){Vr{\v s}nak}, {Ru{\v z}djak}, \&
  {Rompolt}}]{vrsnak91}
{Vr{\v s}nak}, B., {Ru{\v z}djak}, V., \& {Rompolt}, B. 1991, \solphys, 136,
  151

\bibitem[{{Williams} {et~al.}(2005){Williams}, {T{\"o}r{\"o}k}, {D{\'e}moulin},
  {van Driel-Gesztelyi}, \& {Kliem}}]{williams05}
{Williams}, D.~R., {T{\"o}r{\"o}k}, T., {D{\'e}moulin}, P., {van
  Driel-Gesztelyi}, L., \& {Kliem}, B. 2005, \apjl, 628, L163

\end{thebibliography}

% Figures

\newpage

%==========================================================================
%==========================================================================

%==========================================================================
%==========================================================================

%==========================================================================
%==========================================================================

%==========================================================================
%==========================================================================

\end{document}